\documentclass[twocolumn]{article}

\usepackage{times}
\usepackage{appendix}
\usepackage{bm}
\usepackage{graphicx}
\usepackage{amsbsy}
\usepackage{amsmath}
\usepackage{amsfonts}
\usepackage{amsthm}

\usepackage[linkbordercolor={0 0 1}]{hyperref}

\usepackage[all]{hypcap}
\usepackage{tikz}
\usepackage{pgfplots}
\usepackage{verbatim}
\usepackage{enumitem}
\usepackage[linesnumbered, ruled, vlined]{algorithm2e}
\usepackage{float}

\input{Qcircuit}

\usepackage{titling}
\usepackage{authblk}

 \title{ Dynamic Grover Search: \\
   Applications in Recommendation systems and Optimization problems }
 \author {Indranil Chakrabarty, Shahzor Khan and Vanshdeep Singh \textsuperscript{1} }
 \affil{Centre of Security Theory and Algorithmic Research, IIIT-Hyderabad, Hyderabad-500032 Telangana, India}

\begin{document}

\theoremstyle{plain}
\newtheorem{theorem}{Theorem}
\newtheorem{lemma}[theorem]{Lemma}
\newtheorem{corollary}[theorem]{Corollary}
\newtheorem{proposition}[theorem]{Proposition}
\newtheorem{conjecture}[theorem]{Conjecture}

\theoremstyle{definition}
\newtheorem{definition}[theorem]{Definition}


\date{}

\maketitle

\footnotetext[1]{Author order is alphabetic here.}

\begin{abstract} 

\noindent In the recent years we have seen that Grover search algorithm \cite{grover-search} by using quantum parallelism has revolutionized the field  of solving huge class of NP problems in comparisons to  classical systems. In this work we explore the idea of extending Grover search algorithm to approximate algorithms. Here we try to analyze the applicability of Grover search to process an unstructured database with a dynamic selection function in contrast to the static selection function used in the original work \cite{grover-search}. We show that this alteration  facilitates us to extend the application of Grover search to the field of randomized search algorithms. Further, we use the Dynamic Grover search algorithm to define the goals for a recommendation system based on which we propose a recommendation algorithm which uses binomial similarity distribution space giving us a quadratic speedup over traditional classical unstructured recommendation systems. Finally, we see how Dynamic Grover search can be used to tackle a wide range of optimization problems where we improve complexity over existing optimization algorithms.

\end{abstract}

\section{Introduction} \noindent The promise of quantum computation is to enable new algorithms which render physical problems using 
exorbitant physical resources for their solution on a classical computer. There are two broader class of algorithms of which the first class is build upon \textit{Shor's quantum fourier transform} \cite{shor} and includes remarkable algorithms for solving the discrete logarithm  problems providing a astonishing exponential speedup over the best known classical algorithms. The second class of algorithm is based upon Grover's search algorithm for performing \textit{quantum searching} \cite{grover-search}. Apart from these two broader line of divisions Deutsch algorithm based on \textit{quantum parallelism/interference} \cite{Deutsch} is another example which has no classical 
analogue. These algorithms have facilitated us with an unprecedented speed up over the best possible classical algorithms. With the introduction of quantum  algorithms questions were raised for proving
complexity superiority of Quantum Model over Classical Model \cite{feynman}.\\  

\noindent Grover's search algorithm was one of the first algorithms which pioneered a class of problems solvable by quantum computation \cite{grover-framework} facilitating a quadratic speedup over classical systems. Classical unstructured search or processing of search space is essentially linear as we have to process each item using randomized search functions which can be at best optimized to $N/2$ states. In 1996, L.K.Grover gave the Grover Search algorithm to search through a search space in $ \mathcal{O}(\sqrt{N})$ \cite{grover-search}. The algorithm leverages the computational power of superimposed quantum states. In its initialization step an equiprobable superposition state is prepared from the entire search space. In each iteration of this algorithm the coefficients (also known as probability amplitudes) of selected states, based on a selection function, are increased and that of the unselected states are decreased by inversion about the mean. This method increases the coefficients of selected states quadratically and in $ \mathcal{O}(\sqrt{N})$ steps we get the selected states with high probability. The unstructured search approach can be used to compute any NP problem by iterating over the search space.\\

\noindent From a practical perspective quantum search algorithms have several applications such as, it can be used to extract statistics such as the minimal element from an unordered data set more quickly than is possible on a classical computer \cite{min}. It has been extended to solve various problems like finding different measures of central tendency like mean \cite{mean}  and median \cite{median}. It can be used 
to speed up algorithms for NP problems, specifically those problems for which a straightforward search for a solution 
is the best algorithm known. Finally it can be used to speedup the search for keys to the cryptographic systems such as widely used 
Data Encryption Standard (DES).\\

\noindent In the field of e-commerce we have seen recommendation system collects information on the preferences of users for a set 
of items. The information can be acquired explicitly (by collecting user's ratings) or implicitly (monitoring user's behavior) \cite{lee,nun,choi}. 
It make uses of different sources of information for providing user with prediction and recommended items. Further, it tries to balance various factors like accuracy, novelty, dispersity and stability in the recommended items. Collaborative filtering plays an important role in the recommendations although they are often used with other filtering techniques like content-based, knowledge-based. Another important approach in recommending process is the k-nearest neighbor approach in which we find the k-nearest neighbors of the search item. Recently recommendation system implementations has increased and has facilitated in diverse areas \cite{park} like recommending various topics like 
music, television , book documents; in e-learning and e-commerce; application in markets and web search \cite{car,car2,serr,zai,huang,castro,costa,Mcnally}. Mostly these recommendations are 
done in structured classical database .\\

\noindent  NP problems \cite{NPProblems} have been explored in general to be solved using Grover Search \cite{grover-framework}. In extension to that optimization problems have been used to find solution to various specific applications. The class of NP Optimization problem (NPO) \cite{NPO} exists which finds solution for the combinatorial optimization problems under specific conditions.\\


\noindent In this work we develop an extension of Grover's search algorithm by replacing the (static) selection function with a dynamic selection function. This allows us to extend the application of Grover's search to the field of randomized search algorithms, where recommendation systems are one such application. In later sections we define the goals for a recommendation system and propose an algorithm for binomial similarity distribution space giving us a quadratic speedup over traditional unstructured recommendation systems. Another application is in finding an optimal search state for a given NPO problem. We see that Durr and Hoyer's work \cite{min} also performs optimization in $O(\log (N) \sqrt{N})$, however use of dynamic grover search can achieve the same in $O(\sqrt{N})$.\\

\noindent In section II we give a brief introduction of Grover's search algorithm by using standard static selection function. In section III we introduce our model of dynamic Grover's search by defining the algorithm over a binomial distribution space and then by comparing it with the traditional unstructured recommended systems. Lastly, in section IV we provide an application of this dynamic Grover's search in recommendation systems and optimization algorithms.\\

\section{Grover Search Algorithm}
\noindent In this section we briefly describe Grover search algorithm as a standard searching procedure and elaborate on the fact that how it expedites the searching process in contrast to a classical search algorithm on an unstructured database \cite{nielsen}.\\ 

\noindent \textbf{Oracle:} Suppose we wish to search for a given element through an unstructured search space consisting of $N$ elements. For the sake of simplicity, instead of directly searching a given element we assign indices to each of these elements which are just numbers in the range of $0$ to $N-1$. Without loss of generality we assume $N=2^n (n \in \mathbb{Z}^+)$ and we also assume that there are exactly $M$ solutions ($1\leq M \leq N$) to this search problem. Further, we define a selection function $f$ which takes an input state $\ket{x}$, where the index $x$ lies in the range $0$ to $N-1$. It assigns value $1$ when the state is a solution to the search problem and value $0$ otherwise,

\begin{eqnarray}\label{eq:normal_grover}
 f =
  \begin{cases}
   0 & \text{if $\ket{x}$ is not selected}, \\
   1 & \text{if $\ket{x}$ is selected}.
  \end{cases}
\end{eqnarray}

\noindent Here we are provided with quantum oracle-black box which precisely is a unitary operator $O$ and its action on the computational basis is given by,
\begin{equation}
    \ket{x}\ket{q} \xrightarrow{O}  \ket{x} \ket{q \oplus f(x)}.
\end{equation}
\noindent In the above equation we have  $\ket{x}$ as the index register. The symbol $\oplus$ denotes addition modulo $2$ and the oracle qubit $\ket{q}$ gets flipped if we have $f(x)=1$ otherwise it remains  unchanged. This helps us to check whether $\ket{x}$ is a solution to the search problem or not as this is equivalent of checking the oracle qubit is flipped or not.\\


\noindent \textbf{Algorithm:} We start by creating a superposition of $N$ quantum states by applying a Hadamard transformation ($H^{\otimes n}$ where H=$\frac{1}{\sqrt{2}}\bigl(\begin{smallmatrix}
1&1 \\ 1&-1
\end{smallmatrix} \bigr)$) on $\ket{0}^{\otimes n}$.

\begin{equation}\label{eq:grover_input_state}
\ket{\psi} =\frac{1}{\sqrt{N}}\sum _{x=0}^{N-1}\ket{x}  
\end{equation}

\noindent After the initial superposition is ready the algorithm proceeds by repeated application of a quantum subroutine known as the \textit{Grover iteration} or as the Grover operator denoted by $G$. The \textit{Grover iteration} consists of following steps:\\

\SetAlgorithmName{Procedure}{procedure}{List of Procedure}
\begin{algorithm}
  \DontPrintSemicolon
  \Begin{
    Apply the Oracle $O$\;
    Apply the Hadamard transform $H^{\otimes n}$\;
    Perform inversion about mean ($m$) (i.e. apply $ O_{m} = 2 \ket{\psi} \langle \psi| -I $ where $I$ is identity)\;
    Apply the hadamard transform $H^{\otimes n}$\;
  }
  \caption{\textit{Grover iteration}}
\end{algorithm}
\SetAlgorithmName{Algorithm}{algorithm}{List of Algorithm}

\setcounter{algocf}{0}
\begin{algorithm}
\DontPrintSemicolon
\Begin{
  Initialize the system such that its state is given by eq.~\ref{eq:grover_input_state}\;
  Apply \textit{Grover Subroutine} $O(\sqrt{N/M})$ times\;
  Sample the resulting state where we get the expected state with probability greater than $\frac{1}{2}$\;
}
\caption{Grover Search \label{IR}}
\end{algorithm}

\begin{figure}[H]
\[
\Qcircuit @C=1em @R=1em {
& & & & &  &  & \mbox{$O(\sqrt{N/M})$ times } &  & \\
& \lstick{\ket{0}} & {/} \qw & \ustick{n} \qw & \gate{H^{\otimes n}} & \multigate{1}{O} & \gate{H^{\otimes n}} & \gate{O_m} & \gate{H^{\otimes n}} & \qw \\
& \lstick{\ket{1}} & \qw & \qw & \gate{H} & \ghost{O} \gategroup{2}{6}{3}{9}{.8em}{--} & \qw & \qw & \qw & \qw
}
\]
\caption{Circuit for Grover's search}
\end{figure}
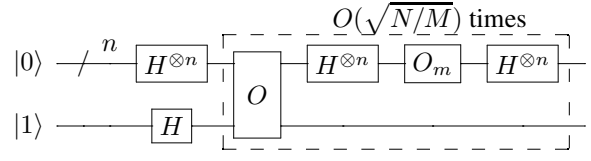



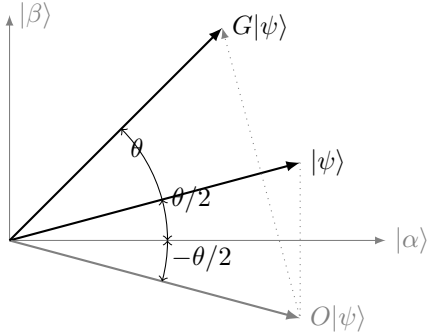
\begin{figure}
\begin{center}
\begin{tikzpicture}
\coordinate (Origin)   at (0,0);
\coordinate (XAxisMax) at (5,0);
\coordinate (YAxisMax) at (0,3);
\draw [thin, gray,-latex] (Origin) -- (XAxisMax) node [right] {$\ket{ \alpha }$};
\draw [thin, gray,-latex] (Origin) -- (YAxisMax) node [right] {$\ket{ \beta }$};
\draw [thick, black,-latex] (Origin) -- (15: 4) node [right] {$\ket{ \psi }$};
\draw [thick, gray,-latex] (Origin) -- (-15: 4)  node [right] {$O \ket{ \psi }$};
\draw [thick, black,-latex] (Origin) -- (45: 4)  node [right] {$G \ket{ \psi }$};
\draw [dotted, gray,-latex] (15: 4) -- (-15: 4) -- (45: 4);

\draw [<->] (0:2.1)  arc (0:15:2.1) node [ right ] {$\theta /2$};
\draw [<->] (0:2.1)  arc (0:-15:2.1) node [above right] {$-\theta /2$};
\draw [->] (15:2.1)  arc (15:45:2.1) node [below right] {$\theta$};
\end{tikzpicture}
\caption{The geometrical representation of single Grover iteration}
\label{fig:grover}
\end{center}
\end{figure}

\noindent \textbf{Geometry:} The process of \textit{Grover iteration} can be decomposed into a two step process i.e. phase rotation of the marked state and a phase rotation about the average which can be expressed geometrically~\cite{geometric-grover-search}, figure~\ref{fig:grover} shows these two phase rotations expressed in a two dimensional space where one dimension represents the solution space and the other represents the remaining search space. These normalized states are written as,

\begin{eqnarray}
\ket{ \alpha} = \frac{1}{\sqrt{N-M}}\sum_x^{''} \ket{ x} \nonumber\\
\ket{ \beta} = \frac{1}{\sqrt{M}}\sum_x^{'} \ket{ x} .
\end{eqnarray}

\noindent The initial state $\ket{ \psi} $ can be re-expressed as 

\begin{equation}
\ket{ \psi} =\sqrt{\frac{N-M}{N}}\ket{ \alpha} +\sqrt{\frac{M}{N}}\ket{ \beta}  .
\end{equation}

\noindent
where $\ket{\alpha}$ represents superposition of all non-solution states and $\ket{\beta}$ represents superposition over all solution states. Geometric visualization of Grover iteration can be described in two parts, the oracle function and inversion about mean. The oracle function can be considered as reflection of state $\ket{\psi}$ about $\ket{\alpha}$ and inversion about mean further reflects this state about the new mean ($\approx \ket{\psi} $). \\

\noindent In short, $G$ can be understood as a rotation in two dimensional space spanned by $\ket{ \alpha} $ and $\ket{ \beta} $. It is rotating the state $\ket{\psi}$ with some angle $\theta$ radians per application of $G$. Applying Grover rotation operator multiple times brings the state vector very close to $\ket{ \beta} $.\\

\noindent \textbf{Success} : The state space after $n$ \textit{Grover iterations} is given as follows,

\begin{equation}
G^n \ket{\psi} = \cos((2n+1)\theta) \ket{\alpha} + \sin((2n+1)\theta) \ket{\beta},
\end{equation}
The success probability ($P_s$) of measuring the solution state $\ket{\beta}$ is $\sin^{2}((2n+1)\theta)$. So, for $P_s = 1$ we need $(2n+1)\theta = \frac{\pi}{2}$ implying $n=\frac{\pi}{4\theta} - \frac{1}{2}$. It is clear that $\frac{\pi}{4\theta} - \frac{1}{2}$ might not always be an integer, so the most optimal strategy would be to choose $n$ such that $\frac{\pi}{4\theta} - \frac{1}{2}$ is as close to $\frac{\pi}{2}$ to maximize $P_s$. This implies that $P_s$ can be very close to 1 but not exactly 1 which indicates a certain failure rate for Grover's search. This gap in Grover's search success has been covered by some extended general algorithms \cite{long-failure-free-grover-search} \cite{exact-grover-search} which propose modifications to phase inversion or introduce arbitrary phases to make the search process exact. \\

\noindent Furthermore, Long et al. extended Grover's search algorithm \cite{long-grover-search} \cite{long-grover-search-2} by using phase matching for quantum searching wherein inversion of marked states is replaced by arbitrary phase rotation $\theta$ and inversion of this prepared state is replaced by rotation through a phase $\phi$. This idea of phase matching has been used by Li et al. to propose a generalized algorithm \cite{phase-matching-grover-search} which demonstrates greater probability of getting correct results when the number of marked states are high ($>\frac{1}{3}$). Alternative approaches have been developed for maximizing the success probability in scenarios where the fraction of  marked states are not known beforehand. Fixed point algorithms \cite{fixed-point-grover} is one such approach which tries to solve the same by always amplifying the marked states. But these algorithms suffer from a major downside i.e. in process of achieving a greater success probability they loose quadratic speedup which the original Grover's search achieved. This loss was eliminated by Theodore et al. at MIT \cite{fixed-point-mit} by presenting an alternative fixed-point algorithm which retains the quadratic speedup while keeping the success probability intact. \\

\noindent \textbf{Entanglement} : In recent years researchers have tried to apprehend the role of entanglement in Grover's search algorithm. It has been understood that entanglement varies with the number of iterations wherein we start with a product state which has zero entanglement and as the algorithm proceeds it reaches a maximum value after which it goes back to zero \cite{grover_entangle} \cite{grover_entangle_pati}.\\

\noindent We use the expression given by eq~\ref{eq:entangle} for calculating the amount of entanglement at each iteration of Grover's search algorithm \cite{grover_entangle}. The equation describes a general geometric measure of entanglement i.e. for a given Grover's state $\ket{\psi_r}$,

\begin{equation}
\ket{\psi_r} = G^r \ket{\psi} = \frac{\cos\theta_r}{\sqrt{2^n - M}} \ket{\alpha} + \frac{\sin\theta_r}{\sqrt{M}} \ket{\beta},
\end{equation}

\noindent where $\theta_r = (r + \frac{1}{2}) \sin^{-1} (2 \sqrt{\frac{M}{N} })$, we calculate its overlap with the nearest $n$-separable state. A general $n$-seperable state is formed by tensor product of $n$ qubits where each qubit is in its most generalized state,

\begin{equation}
  \ket{\xi} = (\cos \frac{\phi}{2} \ket{0} + e^{i\gamma} \sin \frac{\phi}{2} \ket{1} )^{\otimes n},
\end{equation}
where $\phi$ is the azimuthal angle and $\gamma$ is the phase angle. The entanglement for the grover state $\ket{\psi_r}$ can now be expressed as the maximum overlap of between $\ket{\psi_r}$ and $\ket{\xi}$,

\begin{equation} \label{eq:grover_max_entangle}
  E(\ket{\psi_r})  = 1 - max | \langle \xi \ket{\psi_r} |^2 .
\end{equation}

\noindent So, if $\ket{\psi_r}$ is a separable state, which is the case when we start with Grover's search, then the nearest separable state would be the  $\ket{\psi_r}$ itself. Hence, the value of overlap will be $1$ and entanglement will be zero. But in case $\ket{\psi_r}$ is not a separable state then the overlap with the nearest separable state would be less than $1$ and we would have non-zero entanglement.\\

\noindent On simplifying eq~\ref{eq:grover_max_entangle}  \cite{grover_entangle},

\begin{align}\label{eq:entangle}
  E(\ket{\psi_r}) &= 1 - max_{\phi} | \frac{\cos\theta_r}{\sqrt{2^n - M}} ( \cos \frac{\phi}{2} + \sin \frac{\phi}{2} )^n +\nonumber \\
  &(\frac{\sin\theta_r}{\sqrt{M}} +  \frac{\cos\theta_r}{\sqrt{2^n - M}} ) (\sum_{i=1}^{M} \cos^{n - n_i} \frac{\phi}{2} \sin^{n_i} \frac{\phi}{2})
 |^2 
\end{align}

\noindent Using the above expression we can calculate the amount entanglement in Grover's state $\ket{\psi_r}$ after the $r^{th}$ iteration.

\section{Dynamic Grover's Search}
\noindent In this section we introduce a dynamic selection function $f_{s}$ which selects a given state $\ket{ x} $ with certain probability $P_{s}(x)$. We use $f_{s}$ instead of the static selection function $f$ as used in Grover's search to introduce randomness in the search algorithm itself. In principle this  selection criterion can be  based on different properties like similarity to a given state, number of satisfying clauses etc, for applications in recommendation systems, MAX-SAT optimization systems etc. \\

\noindent  We consider $N$ items in a search space which are  represented by (computational) basis vectors in a Hilbert space. Similar to Grover's search we again prepare a superposition of these $N$ items (eq~\ref{eq:grover_input_state}) which acts as an input to our dynamic Grover's search. Here our goal is to select $N_{s}$ states out of these $N$ states using the dynamic  selection function. We define the dynamic selection function  as,

\begin{eqnarray}\label{eq:dynamic_grover}
f_{s} =
\begin{cases}
1 & \text{$\ket{x}$ is selected with $P_{s}(x)$}, \\
0 & \text{otherwise}.
\end{cases}
\end{eqnarray}

\noindent The above selection function is quite similar to the selection function in Grover search (eq~\ref{eq:normal_grover}) in the way that it also gives out $1$ for selected states and $0$ for non selected states. However this function is dynamic in nature because it selects a given state $\ket{x}$ with a probability $P_{s}(x)$ unlike eq~\ref{eq:normal_grover} which always selects the given $\ket{x}$ if it is a part of the solution. This dynamic function is defined based on some selection criteria for a given search scenario and then applied in dynamic Grover's search, for example one could choose fidelity as the selection criteria. It should be noted that in this context this dynamic selection function is predefined and does not change over the course of execution. The dynamic nature of this function introduces selection scenarios that are fundamentally different from the traditional Grover search.\\

\noindent For analysis let us consider the selected states be represented by $\ket{ x_s} $ with coefficient $a_s$ and  unselected states  by $\ket{ x_{us}} $ with coefficients  $a_{us}$. The state of the system at any given time can be represented as,
\begin{equation}
\ket{ \psi } = \sum_{s} a_s\ket{ x_s} + \sum_{us} a_{us}\ket{ x_{us}} . 
\end{equation}

\noindent The probability of sampling from selected states is given by $P_s(=\sum_{s}|a_s|^2)$. Similarly for unselected states we have the corresponding probability as $P_{us}(=\sum_{us}|a_{us}|^2)$. In this context we define a new parameter and call it as gain $G(=\frac{P_s}{P_{us}})$ as an indicator for achieving the desired result. 


\subsection{Analysis of Grovers Search in a different scenario}

\noindent In this subsection we discuss the impact of dynamic selection function on the execution of Grover's search. We also analyze the required conditions for a Grover's step to complete successfully.\\

\noindent \textbf {Lemma 1:} For proper execution of Grover's search following conditions must be satisfied. \\

\begin{enumerate}
\item The mean $\mu$ calculated in the inversion step should be positive. 
\item The probability amplitude $a_{us}$ of the unselected states $\ket{ x_{us}} $ remain positive. 
\item The number of selected states $\ket{ x_{s}} $  for Gain $G$ should be 
\begin{equation}
N_{s} < \frac{N}{2G} \;\;\;\; where \;\; G \gg 1     
\end{equation} 
\end{enumerate}

\noindent \textit{Proof 1:} The mean $\mu_i$ (calculated in the inversion step of $i^{th}$ iteration) should be positive. If the mean is less than 0 then the coefficients of the selected states will decrease and the coefficients of the unselected states will become negative as given by, 

\begin{eqnarray}
&&a'_{x_{s}} = \mu_i - (- a_{x_{s}} - \mu_i) = 2\mu_i + a_{x_{s}}{}\nonumber
\\&&
a'_{x_{us}} = \mu_i - (a_{x_{us}} - \mu_i) = 2\mu_i - a_{x_{us}}
\end{eqnarray}
\hfill $\qedsymbol$

\noindent where $a_{x_{s}}$ is the amplitude of states which are selected in the current iteration $i$ and $a'_{x_{s}}$ is the final amplitude after inversion about mean. Similarly $a_{x_{us}}$ represents the initial amplitude for unselected states and $a'_{x_{us}}$ represents the amplitude for these states after the iteration is complete. \\

\noindent \textit{Proof 2:} The coefficients of the unselected states should remain positive because the mean will be negative in case the coefficient of the unselected state is negative. \hfill $\qedsymbol$\\

\noindent \textit{Proof 3:}
For successfully having a gain $G$, 
\begin{equation}
N_{sel} < \frac{N}{2G} \;\;\;\; where \;\; G \gg 1    
\end{equation}
as described in Appendix A1. \hfill $\qedsymbol$\\

\noindent \textbf{Lemma 2:}
If a state is selected in the current Grover's iteration and not in the next one then the coefficient of this selected state after the next iteration will be less than those states which were not selected in either of these two iterations.\\

\noindent \textit{Proof :} Consider $\ket{ \psi_{i}} $ to be  the input state to the iteration process. The first Grover's step inverts the selected state to $\ket{ \psi_{iv}} $ and then calculates the mean state $\ket{ \psi_{\mu}} $. The final probability of the selected state $\ket{ \psi_{s}} $ is increased and the state $\ket{ \psi_{us}} $ is decreased as compared to the $\ket{\psi_{i}}$, as described earlier in introduction. \\

\noindent Let $a_s$, $a_{us}$ be the coefficients of selected and unselected states respectively, $\mu_1$ be their mean and $a_{s1}$, $a_{us1}$ be the final coefficients after a given Grover iteration. Then we have,\\
\begin{eqnarray}
&& a_{s1} = 2 \mu_1 + a_s {}\nonumber \\&&
 a_{us1} = 2 \mu_1 - a_{us}
\end{eqnarray}

\noindent For the second iteration when no states are selected, let $\mu_2$ be the mean and $a_{s2}$, $a_{us2}$
be the outputs of the iteration. These coefficients are given by,\\

\begin{eqnarray}
&& a_{s2} = 2 \mu_2 - a_{s1} = 2\mu_2 - 2\mu_1 - a_s {}\nonumber \\&&
 a_{us2} = 2 \mu_2 - a_{us1} = 2\mu_2 - 2\mu_1 + a_{us} 
\end{eqnarray}

\noindent Hence, the coefficient of the state $a_{s2}$ that was selected in one iteration is less that the state $a_{us2}$ that was never selected. \\

\begin{figure}[h]
\begin{center}
\begin{tikzpicture}
\coordinate (Origin)   at (0,0);
\coordinate (XAxisMax) at (5,0);
\coordinate (YAxisMax) at (0,3);
\draw [thin, gray,-latex] (Origin) -- (XAxisMax) node [right] {$\ket{\alpha}$};
\draw [thin, gray,-latex] (Origin) -- (YAxisMax) node [right] {$\ket{\beta}$};
\draw [thick, black,-latex] (Origin) -- (15: 4) node [above right] {$\ket{\psi_{i}}$};
\draw [thick, gray,-latex] (Origin) -- (-15: 4)  node [above right] {$ \ket{\psi_{iv}}$};
\draw [dotted, black,-latex] (Origin) -- (13: 4)  node [right] {$ \ket{\psi_{\mu}}$};
\draw [thick, red,-latex] (Origin) -- (11: 4) node [below right] {$ \ket{\psi_{us}}$};
\draw [thick, red,-latex] (Origin) -- (41: 4)  node [above right] {$ \ket{\psi_{s}}$};
\draw [<-] (0:1.5)  arc (0:15:1.5) node [below right] {$\theta$};
\draw [->] (0:1.5)  arc (0:-15:1.5) node [above right] {$-\theta$};
\draw [->] (-15:2.1)  arc (-15:13:2.1) node [below right] {$\theta_{s}$};
\draw [->] (13:2.1)  arc (13:41:2.1) node [right] {$\theta_{s}$};
\draw [->] (15:2.5)  arc (15:11:2.5) node [above] {$2 \theta_{us}$};
\end{tikzpicture}
\caption{Geometrical Representation of selection in Grover Search.
 $\bra{\psi_{i}} \alpha\rangle = - \bra{\psi_{iv}} \alpha\rangle = \cos\theta $,
$\bra{\psi_{iv}} \psi_{\mu}\rangle = \bra{\psi_{s}} \psi_{\mu}\rangle = \cos\theta_s $,
$\bra{\psi_{i}} \psi_{us}\rangle = \cos 2\theta_{us} $  }
\label{fig:GeometricReper}
\end{center}
\end{figure}
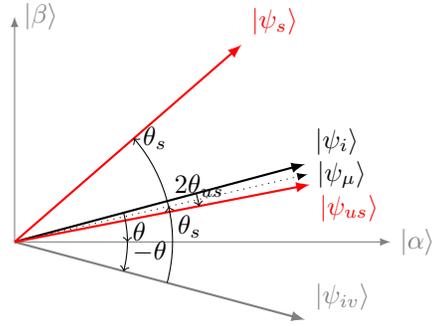

\noindent The Fig~\ref{fig:GeometricReper} shows a geometric representation of a Grover iteration with initial state (black arrow), selected and unselected states (red arrow).\\

\begin{figure}[h]
\begin{center}
\begin{tikzpicture}
\coordinate (Origin)   at (0,0);
\coordinate (XAxisMax) at (5,0);
\coordinate (YAxisMax) at (0,3);
\draw [thin, gray,-latex] (Origin) -- (XAxisMax) node [below right] {$\ket{ \alpha }$};
\draw [thin, gray,-latex] (Origin) -- (YAxisMax) node [right] {$\ket{ \beta }$};
\draw [thick, black,-latex] (Origin) -- (13: 4) node [below right] {$ \ket{ \psi_{us1} }$};
\draw [thick, black,-latex] (Origin) -- (27: 4)  node [above right] {$ \ket{ \psi_{s1} }$};
\draw [dotted, black,-latex] (Origin) -- (15: 4)  node [right]  {$ \ket{ \psi_{\mu2} }$};
\draw [thick, red,-latex] (Origin) -- (17: 4) node [above right] {$\ket{ \psi_{us2} }$};
\draw [thick, red,-latex] (Origin) -- (3: 4)  node [right] {$ \ket{ \psi_{s2} }$};

\draw [->] (27:1.5)  arc (27:3:1.5) node [below right] {$2 \theta_{s2}$};
\draw [<-] (17:2.5)  arc (17:13:2.5) node [below right] {$2 \theta_{us2}$};
\end{tikzpicture}
\caption{Geometrical representation of the rejection of a selected state. Here $\bra{\psi_{us2}} \psi_{us1}\rangle = \cos\theta_{us2} $, $\bra{\psi_{s1}} \psi_{s2}\rangle = \cos\theta_{s2} $ }
\label{fig:GeometricReper2}
\end{center}
\end{figure}
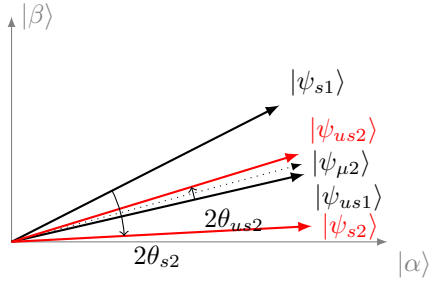

\noindent Further, let us suppose that in the next iteration a state $\ket{ \psi_{s}} $ which was previously selected gets unselected and another state $\ket{ \psi_{us}} $ which was not selected previously is still not selected.
Since no state is selected in this iteration no inversion occurs, so the mean $\ket{ \psi_{\mu2}} $ will lie above $\ket{ \psi_{us}} $ as shown in Fig~\ref{fig:GeometricReper2}.  The inversion about mean will cause the state $\ket{ \psi_{s1}} $ (in black) to become the state $\ket{ \psi_{s2}} $ (in red). Similarly state $\ket{ \psi_{us1}} $ (in black) will become state $\ket{ \psi_{us2}} $ (in red). It is quite evident that now $\ket{ \psi_{s2}} $ has a probability lower than those states $\ket{ \psi_{us2}} $ which were never selected in either step.

\noindent Note that the issue is not restricted to the case where no state is selected, it is inherent with the use of inversion function for unselected states. In order to overcome this issue we need to always run each Grover's iteration twice with a given result from the selection function $f_s$ so that the relative coefficients remain in the order of number of times the state was selected.\\

\noindent \textbf{Lemma 3:} If no state is selected and the Grover's step is repeated twice no change happens in the coefficients.\\

\noindent \textbf{Lemma 4:} If all states are selected and the Grover's step is repeated twice, there is no net change in the coefficients.\\

\noindent \textit{Proof:} In first step all the coefficients will become negative of themselves so the mean will be negative and inversion around the mean will leave them in negative coefficients. In second step all the coefficients will become negative of themselves so mean will be positive and the coefficients will be rotated back to their original places around the mean. \\

\subsection{Dynamic Grover's Search Algorithm}

\noindent Algorithm~\ref{algo:dgs} gives the formalized procedure of dynamic Grover's search algorithm.\\

\SetAlgorithmName{Procedure}{procedure}{List of Procedure}
\setcounter{algocf}{1}
\begin{algorithm}
\DontPrintSemicolon
\Begin{
  Apply the Oracle $f_{s}$ and store the result\;
  Apply the \textit{Grover Iteration} using the stored oracle results\;
  Apply the \textit{Grover Iteration} again using the stored oracle results, to nullify any negative affects of inversion about the mean.\;
}
\caption{\textit{Dynamic Grover Iteration}} 
\end{algorithm}
\SetAlgorithmName{Algorithm}{algorithm}{List of Algorithm}
\setcounter{algocf}{1}



\begin{algorithm}
\DontPrintSemicolon
\Begin{
  Initialize the system, such that there is same amplitude for all the N states\;
  Apply \textit{Dynamic Grover Iteration} $O(\sqrt{N})$ times\;
  Sample the resulting state, where we get the expected state with $probablity > \frac{1}{2}$\;
}
\caption{Dynamic Grover Search}\label{algo:dgs}
\end{algorithm}


\section{Applications of Dynamic Grover's search}
\noindent In this section we explore two different applications of our dynamic Grover's search algorithm. In the first subsection we demonstrate its application in developing a recommendation system and in the second section we describe a generic optimization problem and discuss its solution using dynamic Grover's search.

\subsection{Quantum Recommendation Algorithm}\label{sec:qreco}

\noindent A recommendation algorithm deals with the problem of finding similar items to a given item, for example consider a user who is shopping an e-commerce website for a product, the task of a recommendation algorithm is to suggest products which might be similar to the former. The similarity can be decided on various different parameters so as to achieve meaningful results which in turn may help in boosting sales. Although, state of the art recommendation algorithms are quite complex one can think of them as similarity searching procedures. In essence a recommendation algorithm  on an unstructured search space is similar to Grover's search with the difference of selection function. If we know the search space well we can construct a static selection function which can select top $M$ states. In case of unknown search space or a dynamic search space, we may not always be able to construct a static selection function. In that case we need to associate the selection dynamically to the similarity with the given state $\ket{x} $ (say). This will increase the probability of selecting the desired $M$ states sufficiently.\\

\begin{enumerate}[leftmargin=*]

\item \textbf{Recommendation Problem}:

Consider a standard recommendation problem,
\begin{itemize} 
\item Given a unstructured search space $S$, 
\item The dimensionality of the space be $n$,  and total number of
states to be N($=2^n$).
\item We need to find $M$ recommended states for a given search result $\ket{x} $.
\end{itemize}

\noindent Let the similarity $S(x,y)$ of two pure states $\ket{x} ,\ket{y} $ represent a measure of the likeliness of these two states to be recommended for each other.\\

\pgfmathdeclarefunction{gauss}{2}{%
  \pgfmathparse{1/(#2*sqrt(2*pi))*exp(-((x-#1)^2)/(2*#2^2))}%
}
\begin{figure}[h]
\centering
\begin{tikzpicture}
\begin{axis}[
  no markers, domain=0.8:7.5, samples=100,
  axis lines*=left,xlabel=$x$, ylabel=$N_{States}$,
  every axis y label/.style={at=(current axis.above origin),anchor=south},
  every axis x label/.style={at=(current axis.right of origin),anchor=west},
  height=5cm, width=8cm,
  xtick=\empty, ytick=\empty,
  enlargelimits=false, clip=false, axis on top,
  grid = major
  ]
  \addplot [thick,cyan!50!black] {gauss(4,1)};
  \draw [yshift=-0.6cm, xshift=-1.3cm, latex-latex](axis cs:4,0) -- node [fill=white] {$S(x,y)$} (axis cs:7.2,0);
\end{axis}
\end{tikzpicture}
\caption{Distribution of the states with respect to similarity function (Similarity function $S(x,y)$ vs $N_{States}$)}
\label{fig:BinomialDistribution}

\end{figure}
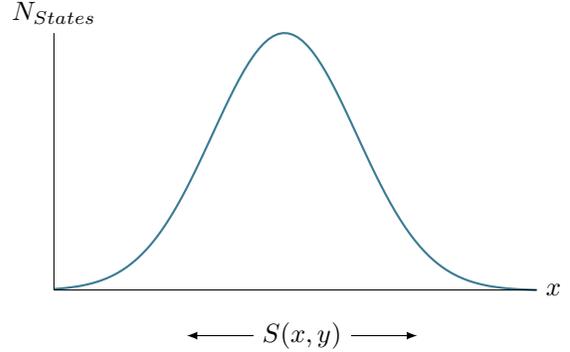

\item \textbf{Criteria for an effective Recommendation function}:
\noindent Now we give a criteria for selection function $f_{s}$ to be effective in a dynamic system. Let the dynamic selection function be given by eq.~\ref{eq:dynamic_grover}. Then $P_{s}$ should have the following criteria in order for the selection function to give good recommendations:
\begin{itemize}
\item The most likely state is selected with high probability,
$\lim_{S(x,y) \to n } P_{s}(x) \geq \Big( 1-\frac{1}{N} \Big)$ i.e. if $x$ and $y$ are similar at $h$ bits where $h \to n$ then probability of selection should be high.
\item The least likely state is selected with a low probability,
$\lim_{S(x,y) \to 0 } P_{s}(x) \leq \Big( \frac{1}{N} \Big)$, i.e. if $x$ and $y$ are similar at $h$ bits where $h \to 0$ then such state should be selected with low probability.
\item In order to select $M$ states, we have from lemma 1, $M < \frac{N}{2G}$. So, the expected number of selected states, $E(N_{s}) = \int_{x} P_{s}(x,y) \approx M$
\end{itemize}

\pgfmathdeclarefunction{selectionProb}{2}{%
	\pgfmathparse{-1/(x*x*x-0.1) +#1}%
}
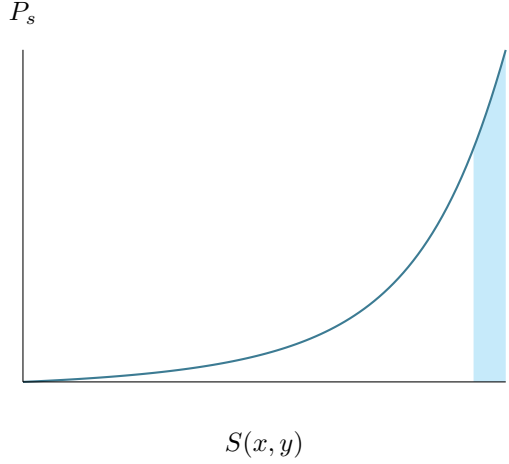
\begin{figure}[h]
\begin{center}
\begin{tikzpicture}
\begin{axis}[
  no markers, domain=0.8:7.5, samples=100,
  every axis y label/.style={at={(ticklabel* cs:1.05)},anchor=south},
  axis lines*=left,xlabel={$S(x,y)$}, ylabel={$P_s$},
  height=6cm, width=8cm,
  xtick=\empty, ytick=\empty,
  enlargelimits=false, clip=false, axis on top,
  grid = major
]
\addplot [fill=cyan!20, draw=none, domain=-0.6:-0.5] {selectionProb(0,10)} \closedcycle;
\addplot [thick,cyan!50!black,domain=-2:-0.5] {selectionProb(0,10)};
\end{axis}
\end{tikzpicture}
\caption{Plotting probability of selection $P_s$ with the similarity function $S(x,y)$. The blue shaded region indicates the expected selected items.}
\label{fig:BinomialProbabilty}
\end{center}
\end{figure}

\item \textbf{Recommendation for Binomial distribution}:
\noindent Consider an example for initial state space with equal initial probability for each of the states. The similarity of states with respect to a particular state is given by the hamming distance (i.e. number of bits which differ between two binary strings) between the states. The similarity function $S(x,y)$ would be a binomial curve Fig~\ref{fig:BinomialDistribution}. The probability of selection is given by the following equation:
\begin{equation}
 e^{-\log(\sqrt[n]{K}-1) S(x,y)}
\end{equation}

\noindent  The figure~\ref{fig:selection_circuit} shows a sample implementation of the selection function $f_s$. This function probabilistically selects states which are similar to the searched state, say $\ket{x_0}$. 
\begin{figure}[H]
\[
\Qcircuit @C=1em @R=.7em {
  & \lstick{\ket{x}}  & \multigate{2}{O_{s, \ket{x_0}}} & \qw & \qw &
  \qw  & \qw  & \rstick{\ket{x}} \\
  & &  &  &  & & & \\
  & \lstick{\ket{0}}  & \ghost{O_{s, \ket{x_0}}}  & \ustick{\ket{s}} \qw  & \qw  &
  \multigate{3}{O_c}  & \qw  & \rstick{\ket{s}}  \\
  & &  &  &  & & & \\
  & \lstick{\ket{\zeta_{ud}}} & \gate{O_r} & \ustick{\ket{r}} \qw  & \qw  & \ghost{O_c} & \qw  & \rstick{\ket{r}}  \\
  & \lstick{\ket{0}} & \qw & \qw  & \qw  & \ghost{O_c} &  \qw  & \rstick{\ket{O_1}}
}     
\]
\caption{Circuit representation of selection function}\label{fig:selection_circuit}
\end{figure}
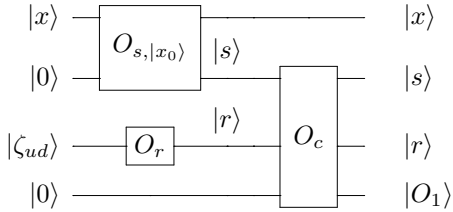

\begin{itemize}
\item $|x\rangle$ : a state in the input superposition of N ($=2^n$) states.
\item $|x_0\rangle$ : searched state.
\item $O_{s, \ket{x_0}}$ : operator for calculating similarity of a given input state $\ket{x}$ to $\ket{x_0}$ \cite{qcircuit-similarity}.
\item $O_c$ : operator for comparing two numbers of $n$ bits \cite{qcircuit-similarity}.
\item $O_r$ : operator for generating random number \cite{qcircuit-random}.
\item $\ket{\zeta_{ud}}$: user defined input for the random number generator $O_r$.
\item $O_1$ : answer bit.

\end{itemize}

\begin{equation}
\begin{split}
O_{s, \ket{x_0}} \ket{x} \ket{0}^{\otimes n} &\to \ket{x} \ket{0 \oplus f_{x0}(x)} \to \\
& \ket{x} \ket{f_{x0}(x)} \to \ket{x} \ket{s}
\end{split}
\end{equation}


where $|s\rangle = |f_{x0}(x)\rangle$ and $f_{x_0}(x)$ gives similarity $s$, of a given input state $\ket{x}$ to the searched state $\ket{x_0}$ such that $ 0 \leq s \leq n $. Also we generate a random number $r$ using oracle $O_r$ such that $ 0 \leq r \leq n $. One could use an equiprobable superposition state i.e. $\ket{\psi_{er}} = \frac{1}{\sqrt{N}} (\ket{0} + \dots + \ket{n})$ as input to a random number generator in which case $O_r$ will be a measurement operator performing a measurement on $\ket{\psi_{er}}$ in computational basis leaving us with a random outcome. We then feed $\ket{s}$, $\ket{r}$ and an answer bit $\ket{0} $ into the circuit $O_c$,

\begin{equation}
 O_c \ket{s} \ket{r} \ket{0} \to \ket{x} \ket{r} \ket{O_1}
\end{equation}


\noindent $O_c$ compares $\ket{s}$ and $\ket{r}$ and if $s \geq r$ then it flips the answer bit to $\ket{1}$ ie. the given input state is selected else it keeps the answer bit unchanged meaning that the given input state is not selected. \\

\noindent The expected selected states should have similarity as shown in Fig~\ref{fig:BinomialProbabilty} i.e. states which are very similar to the searched state should have a higher selection probability.

\end{enumerate}

\subsection{Approximate Optimization Algorithms }

\noindent The Grover's search algorithm is a landmark algorithm because it provided a framework \cite{grover-framework} which can be used to solve any NP problem with a quadratic speedup over classical systems. Durr and Hoyer's work on finding the minimum in a given search space \cite{min} can be considered as a tool to find the optimal value (min or max) for a search space. But it is achieved by applying Grover's search algorithm multiple times and uses the quantum probability during the sampling to arrive at the optimal state.\\

\begin{enumerate}[leftmargin=*]
\item \textit{Optimization Problem}: An optimization problem can be represented in the following way: \\ \\
\noindent  $\boldsymbol{Given:}$ A function $f : A \to R$ from some set $A$ to the set of all real numbers $R$. \\
$\boldsymbol{Sought:}$ An element $\ket{x_o} $ (optimal) in $A$ such that $f(\ket{x_o})  \leq f(\ket{x})$ 
for all $\ket{x} $ in $A$ (minimization) or such that    $ f(\ket{x_o}) \geq f(\ket{x})$ for all $\ket{x} $ in $A$ (maximization). \\


\item \textit{Solving Optimization Problem using Durr and Hoyer's Min Approach}: To solve the Optimization problem 
using Durr and Hoyer's approach \cite{min}, the selection function would use algorithm \ref{IR}.

\begin{algorithm}[h]
\DontPrintSemicolon
\Begin{
  Set $\ket{x_o}  = \ket{0} $\;

  \While{$True$}{
    Set $f_{s} = f(\ket{x} ) \geq f(\ket{x_o} )$\;
    Run Grover Search using $f_{s}$, sample out $\ket{y} $\;
    \eIf{ $ f(\ket{y} ) > f(\ket{x_o}) $}{
      $\ket{x_o}  = \ket{y} $\;
    } {
      break;
    }

  }
  return $\ket{x_o} $\;
}
\caption{Durr and Hoyer’s approach \label{IR}}
\end{algorithm}


\noindent The algorithm runs in expected $O(\log(N)\sqrt{N})$ Grover iterations.\\

\item \noindent With dynamic Grover's search we present a generic framework for solving optimization problems using classical probability. Consider the distribution function $D(\ket{x} ) = f(\ket{x} )$, we use a probabilistic function $P_{s}: A \to [0,1]$ such that,

\begin{equation}
\lim_{f(\ket{x} ) \to f_{max} } P_{s}(\ket{x} ) \geq \Big( 1-\frac{1}{N} \Big)
\end{equation}

\begin{equation}
\lim_{f(x) \to f_{min}} P_{s}(\ket{x} ) \leq \Big( \frac{1}{N} \Big)
\end{equation}

\noindent Using a good heuristic a probabilistic function $P_{s}$ can be chosen and we can get optimal results with high probability by running dynamic Grover's search algorithm. Hence it is apparent that dynamic Grover's search can be modeled for any optimization problem which runs in $O(\sqrt{N})$ Grover's iterations and the accuracy of this search depends on the probability function $P_{s}$.\\

\end{enumerate}

\section{Results and Discussion}
\begin{figure}[h]
\begin{center}
\begin{tikzpicture}[x=1cm,y=0.2cm]
\begin{axis}[
axis lines=left,
axis on top=true,
xmin=0,
xmax=15,
xlabel={$S(|x\rangle, |y\rangle)$},
xticklabel style={/pgf/number format/1000 sep=},
ymin=0,
ymax=1.0,
ylabel={Probability}
]

\addplot[color=blue,mark=o] coordinates {
	( 0 ,   0                 ) 
	( 1 ,   0                 ) 
	( 2 ,   0                 ) 
	( 3 ,   7.31511470083e-06 ) 
	( 4 ,   0.000105024146776 )
	( 5 ,   0.000570578946665 )
	( 6 ,   0.00249286535328  )
	( 7 ,   0.00745920069926  )
	( 8 ,   0.0317610356855   )
	( 9 ,   0.0618686957303   )
	( 10,   0.120915738795    )
	( 11,   0.179278528288    )
	( 12,   0.240100783919    )
	( 13,   0.35544023332     )
};

\addplot[smooth, color=red,mark=o] coordinates {
	
	( 0 ,  0                        )
	( 1 ,  0                        )
	( 2 ,  0                        )
	( 3 ,  6.428343871e-07          )
	( 4 ,  9.22926512908e-06        )
	( 5 ,  5.01410821938e-05        )
	( 6 ,  0.000137933892775        )
	( 7 ,  0.000211768013807        )
	( 8 ,  0.000193217650065        )
	( 9 ,  0.000106848258484        )
	( 10,  3.51263075808e-05        )
	( 11,  6.70384432261e-06        )
	( 12,  0.932631829595           )
	( 13,  0.0666165592568          )
};
\end{axis}
\end{tikzpicture}
\caption{Plotting probability of sampling with the similarity function $S(x,y)$. The (red) line indicates standard Grover's algorithm while the (blue) line indicates our dynamic grover search algorithm}
\label{fig:accuracy}
\end{center}
\end{figure}
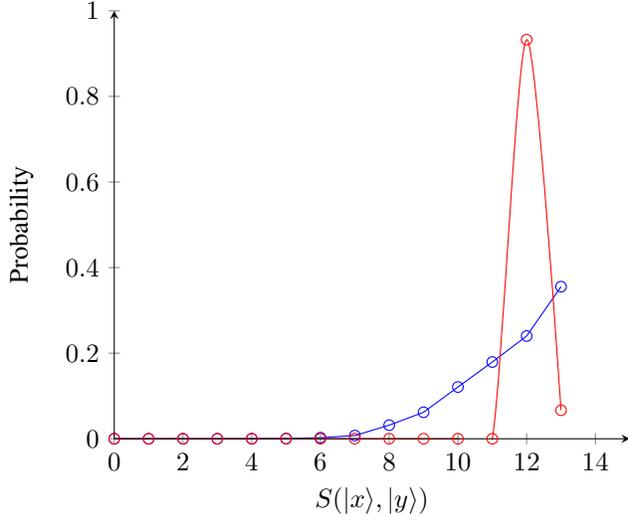

\noindent  On simulating the scenario described in section~\ref{sec:qreco} we see that dynamic Grover's search gives a similar performance (Fig~\ref{fig:ExecutionComparison}) as well as desired accuracy (Fig~\ref{fig:accuracy}) as would have been given by a static selection function using the Grover's search. However this algorithm makes our recommendation system robust with respect to changes in search space and distribution of search space. Further details can be seen from the Appendix.

\begin{figure}[h]
\begin{center}
\begin{tikzpicture}[x=1cm,y=0.2cm]
\begin{axis}[
axis lines=center,
axis on top=true,
xmin=10,
xmax=14,
xlabel={Number of bits},
xticklabel style={/pgf/number format/1000 sep=},
ymin=0,
ymax=70,
ylabel={Number of steps}
]

\addplot[color=blue,mark=o] coordinates {
	( 10,  7  )
	( 11,  11 )
	( 12,  17 )
	( 13,  22 )
	( 14,  32 )
};
\addplot[color=red,mark=o] coordinates {
	( 10,  7  )
	( 11,  10 )
	( 12,  13 )
	( 13,  18 )
	( 14,  25 )
};
\end{axis}
\end{tikzpicture}
\caption{Comparative analysis of the number of steps with the dimensionality of the search space , Dynamic Grover(blue), Grover(red)}
\label{fig:ExecutionComparison}
\end{center}
\end{figure}
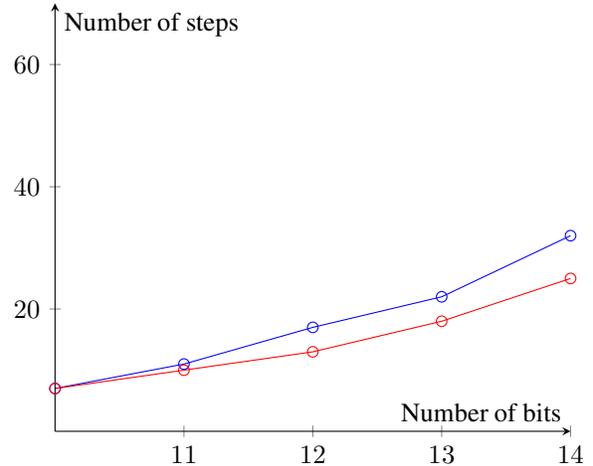


\noindent While analyzing figure~\ref{fig:ExecutionComparison} it must not be thought of as a performance comparison between standard and dynamic Grover's search. It must be duly noted that the aim of dynamic Grover's search is not to outperform standard Grover's search (which has already been proven to be optimal \cite{optimal-grover-search}) but to add flexibility to its searching process. The comparison depicted in figure~\ref{fig:ExecutionComparison} shows that dynamic Grover's search follows a similar trend to that of standard Grover's search. Furthermore, figure~\ref{fig:accuracy} shows the sampling probability for standard and dynamic Grover's search. The plot shows that dynamic Grover's search gives non zero sampling probability for states which are similar to the searched state. But for standard Grover's search we can see only one peak which corresponds to the state which was being searched. Furthermore, due to the fact that dynamic Grover's search tries to increase the sampling probability for more than one states it may take more steps to reach its solution state which explains the results shown by fig~\ref{fig:ExecutionComparison}.

\noindent It is apparent that the dynamic oracle presented in the above discussion has become non-unitary in nature because of the measurement performed in the quantum random number generator which we have introduced in order to induce probabilistic selection of a given input state. Like Grover's search, dynamic grover's search has a complexity of $O(\sqrt{N})$ and in each of these iterations our quantum oracle selects states probabilistically using the quantum random number generator. It must be duly noted that similar to Grover's search oracle which outputs $0$ or $1$ for a given state indicating whether the state is selected or not, the quantum oracle in dynamic Grover's search also outputs $0$ or $1$ but it does so probabilistically. This probabilistic selection is achieved by modifying the original oracle to use a random number generator. \\

\noindent Since dynamic Grover's search has a similar complexity to that of Grover's search which means we would require almost $\sqrt{N}$ random numbers, one for each iteration. These random numbers are fed into the circuit as $\sqrt{N}$ additional quantum states ($\ket{\psi_{rs}}$) each of which will be superposition states,
\begin{equation}\label{eq:equiprobable_superposition_state}
\ket{\psi_{rs}} = \frac{ \ket{0} + \dots + \ket{x} + \dots + \ket{n}}{\sqrt{n}} = \frac{1}{\sqrt{n}} \sum_{x=0}^n \ket{x}
\end{equation}
where $x: 0 \to n$, denotes the similarity between any two given states, with a maximum value of $n$. So if we wish to search a $n$ qubit state in a given search space then the maximum similarity between the searched state and any state from the search space will be $n$ i.e. all the $n$ qubits are same. Note that we can also use superposition states $\ket{\psi_{rs}}$ that are not equiprobable, which will be the case when we wish to search for more useful solutions. \newline
As previously mentioned, we need to supply $\sqrt{N}$ random number generating states $\ket{\psi_{rs}}$ as input along with the (input) search space $\ket{\psi_{G}}$. Hence, the initial state for dynamic Grover's search can be thought of as follows,
\begin{equation}
\ket{\psi_{DG}} = \ket{\psi_{G}} \ket{\psi^{0}_{rs}} \ket{\psi^{1}_{rs}} \dots \ket{\psi^{\sqrt{N}}_{rs}}
\end{equation}
where $\ket{\psi_{G}}$ is search space (grover state),
\begin{equation}
\ket{\psi_{G}} = \frac{ \ket{00\dots0} + \dots + \ket{10\dots1} + \dots + \ket{11\dots1}}{\sqrt{N}}
\end{equation}
and $\ket{\psi^{i}_{rs}}$ is represented by eq~\ref{eq:equiprobable_superposition_state}.\\ \\
In each iteration of dynamic Grover's search we measure one of these additional quantum states $\ket{\psi^{i}_{rs}}$, which gives us a random number which is used for probabilistic selection of states in that iteration. For example, during the first iteration we measure $\ket{\psi^{0}_{rs}}$ which may leave us with the following state,
\begin{equation}
\ket{\psi^0_{DG}} = \ket{\psi_{G}} \ket{x} \ket{\psi^{1}_{rs}} \dots \ket{\psi^{\sqrt{N}}_{rs}}
\end{equation}
where $x$ lies between $0$ and $n$. After this measurement we use the corresponding output i.e. the random number $x$ in our oracle to probabilistically mark states in the input search space. So, states which show similarity greater than $x$ to the searched state are marked which may leave us with the following Grover's state,
\begin{equation}
\ket{\psi^0_{G}} = \frac{ \ket{00\dots0} -\ket{00\dots1} + \dots - \ket{11\dots1}}{\sqrt{N}}
\end{equation}
Now we invert the above state about mean and hence we have the final dynamic Grover's state after the first iteration,
\begin{equation}
\ket{\psi^{1}_{DG}} = \ket{\psi^{1}_{G}} \ket{x} \ket{\psi^{1}_{rs}} \dots \ket{\psi^{\sqrt{N}}_{rs}}
\end{equation}
where $\ket{\psi^{1}_{G}} $ is the grover's state after inversion about mean. \newline
So, it is apparent that even though we perform a measurement in our dynamic oracle the probability conservation for the Grover's state holds because the dynamic Grover's state is a product state and we perform our non-unitary operation on the additional superposition states for the purpose of generating random numbers. However, in a situation where the search space and probabilistic selection space (i.e. the space from which we generate our random numbers by performing measurements) become entangled somehow then the probability gets affected which can be an independent problem of its own. \\

\begin{figure}[h]
\begin{center}
\begin{tikzpicture}[x=1cm,y=0.2cm]
\begin{axis}[
  axis lines=left,
  axis on top=true,
  xmin=0,
  xmax=30,
  xlabel={Number of steps},
  xticklabel style={/pgf/number format/1000 sep=},
  ymin=0,
  ymax=1.0,
  ylabel={$E(\ket{\psi_r})$}
]

\addplot[color=blue] coordinates {
  (1 ,   0.013695269603828941)
  (2 ,   0.09168385604938978 )
  (3 ,   0.15097958539093592 )
  (4 ,   0.21770045659548642 )
  (5 ,   0.46702147179625175 )
  (6 ,   0.5766809316161363  )
  (7 ,   0.5050845712569423  )
  (8 ,   0.8116517806594554  )
  (9 ,   0.8164723742846256  )
  (10,   0.9584051649717154  )
  (11,   0.9352640839967793  )
  (12,   0.9546365955057109  )
  (13,   0.8281101880632697  )
  (14,   0.9355821052992698  )
  (15,   0.6951841163154999  )
  (16,   0.9292576393637849  )
  (17,   0.5331239819085476  )
  (18,   0.5303283872045728  )
  (19,   0.2110886244962381  )
  (20,   0.3517763540008917  )
  (21,   0.2379635282800705  )
  (22,   0.1405279820417451  )
  (23,   0.028527341428792763)

};

\addplot[smooth, color=red] coordinates {
	
  (1 ,   0.012715482577012271)
  (2 ,   0.05017587394628864 )
  (3 ,   0.11047889671672995 )
  (4 ,   0.1905622880132749  )
  (5 ,   0.2863593131216461  )
  (6 ,   0.3930052793605583  )
  (7 ,   0.5050845712569423  )
  (8 ,   0.6169056622691851  )
  (9 ,   0.7227901376193205  )
  (10,   0.817361051293475   )
  (11,   0.8958159740816327  )
  (12,   0.9249323422981167  )
  (13,   0.9232365381469698  )
  (14,   0.923403802871676   )
  (15,   0.9254408326242531  )
  (16,   0.9292576393637849  )
  (17,   0.8831808251891625  )
  (18,   0.8015300518011432  )
  (19,   0.7045672054700975  )
  (20,   0.5972161807362731  )
  (21,   0.4849283968244352  )
  (22,   0.37340596776135593 )
  (23,   0.2683121418978959  )
  (24,   0.17498371507868338 )
  (25,   0.09816002147813141 )
  (26,   0.04174226428689298 )
  (27,   0.008595407737336136)

};
\end{axis}
\end{tikzpicture}
\caption{Plot showing entanglement vs number of steps. The red line indicates standard Grover's search algorithm while the blue line indicates our dynamic grover search algorithm}
\label{fig:Entanglement_Analysis}
\end{center}
\end{figure}
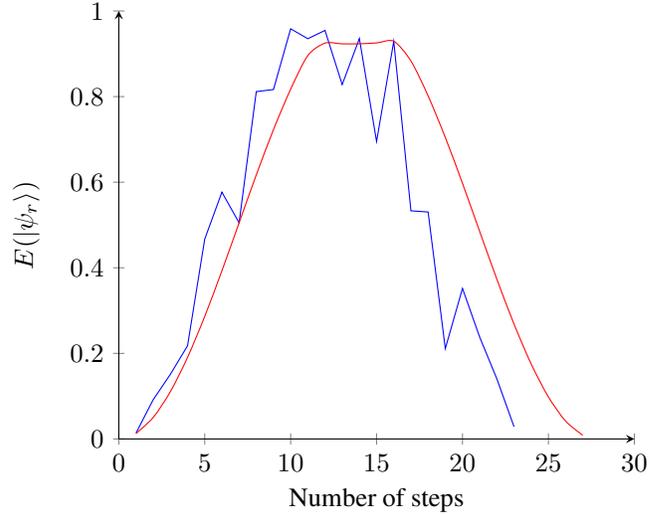

\noindent Further, we use equation \eqref{eq:entangle} and perform entanglement comparison between normal and dynamic Grover's search. In figure~\ref{fig:Entanglement_Analysis} we plot the variation of entanglement with the number of iterations both for Grover's search and dynamic Grover's search algorithm in context of the recommendation problem. 

\noindent From the plot we can observe that entanglement for normal Grover's search increases and decrease smoothly but such is not the case for dynamic Grover's search. This is expected because dynamic Grover's search keeps selecting and deselecting states. Whenever the number selected states is more in comparison to normal Grover's search we see that entanglement is higher but the overall trend of entanglement in dynamic Grover's search is similar to (normal) Grover's search. \\

\noindent \textbf{Convergence}: It is apparent from our discussion that the convergence of dynamic Grovers search is completely dependent on selection function $f_s$. If the selection function is static then the algorithm proceeds similar to normal Grover's search but in case the selection function is probabilistic then convergence of algorithm depends on probabilistic nature of this selection function, for example if the probabilistic function keeps selecting and de-selecting a single state from one iteration to another then algorithm will not be able to converge to any solution. Another example where the algorithm would fail to converge is when the probabilistic selection function selects a lot states in which case the \textit{Grovers iteration} will fail to increase the amplitude of selected states as the mean might become negative. So, not every selection function is good fit for dynamic grover's search and it is not possible to analytically reason about convergence in all such possible dynamic search scenarios. Nevertheless, we suggest guidelines which might help identify suitable probabilistic selection functions. The selection function should be such that it selects few states with high probability which would ensure that in each iteration only those states are selected and the mean remains positive. Such a probabilistic selection function would also ensure that those few states are repeatedly selected in each iteration with high probability thus ensuring that dynamic Grover's search will converge in the same way as normal Grover's search.\\

\section{Conclusion}
In our work we have presented an extension to the current Grover's search algorithm. The extended algorithm or dynamic Grover's search proves very effective in randomized searching, for eg. scenarios wherein the searched item might not exist in the searched space and it is acceptable to output an item which is similar to the searched item. Dynamic Grover's search doesn't offer a speed up over the standard Grover's search as it is optimal but it is worthwhile to mention that dynamic Grover's search allows it's user to choose their own similarity criteria which makes it a general framework for carrying out the task of searching similar items. We have explored the same by proposing a quantum recommendation system and have suggested that similar approach could be used to tackle various different problems.


\section{Appendices}

\begin{appendices}
\section{ Proof of Lemma 1}
	
\noindent	\textbf{Lemma 1.}
	\textit{In order for the Grover's search to have a meaningful next step following conditions must be satisfied.}
	
\begin{enumerate}
    \item The Mean ($\mu$) (calculate in the inversion step) should be positive.
    \item The coefficients of the unselected states should remain positive.
    \item The number of selected states $N_{s}$ for Gain G($=\frac{P_{s}}{P_{us}})$ should be
		\[ N_{s} < \frac{N}{2G} \;\;\;\; where \;\; G \gg 1 \]

\end{enumerate}

\noindent \textbf{Proof:} Let $N$, $N_s$ and $N_{us}$ represent total number of states, number of selected states, and number of unselected states respectively.
\noindent	Hence 
	\begin{equation}
	     N_{us} = N - N_s
	\end{equation}
	
\noindent	Let $\mu$, $a_s$ and $a_{us}$ represent the mean, the coefficient of selected states, and coefficient of unselected states respectively.
	So, 
	\begin{equation}
	\mu = \frac{N_{us}  a_{us} - N_s  a_s}{N}    
	\end{equation}
	
\noindent For coefficient of  unselected states to be positive (say in the last step)
	\begin{equation}
	    a_{us1} = 2\mu - a_{us} > 0	    
	\end{equation}
	\begin{equation}
	    \implies 2\frac{N_{us}  a_{us} - N_{s}  a_{s}}{N} - a_{us} > 0	    
	\end{equation}
    \begin{equation}
	    \implies \frac{a_s}{a_{us}} < (\frac{N}{2 N_s} - 1)        
    \end{equation}
	
\noindent Now $G = \frac{P_{s}}{P_{us}}$,
    \begin{equation}
	    \implies G = \frac{N_s  a_s^2}{(N - N_s)  a_{us}^2}        
    \end{equation}	
	
\noindent since $a_s$ and $a_{us}$ is positive,
	\begin{equation}
	    \implies G < \frac{N_s}{N - N_s}  (\frac{N}{2 N_s} - 1)^2	    
	\end{equation}

\noindent for $G \gg 1 $,
    \begin{equation}
	    N_s < \frac{N}{2 G}        
    \end{equation}

\hfill $\qedsymbol$	 

\end{appendices}

\end{document}